# Trapping and Assembly of Living Colloids at Water/Water Interfaces

Sarah D. Hann,[a] Mark Goulian,[b] Daeyeon Lee,[a,*] and Kathleen J. Stebe[a,*]

We study the assembly of colloids in a two phase water-water system that provides an environment that can sustain bacteria, providing a new structure with rich potential to confine and structure living colloids. The water-water system, formed via phase separation of a casein and xanthan mixture, forms a 3-D structure of coexisting casein-rich and xanthan-rich phases. Fluorescent labelling and confocal microscopy reveal the attachment of these living colloids, including *Escherichia coli* and *Pseudomonas aeruginosa*, at the interface between the two phases. Inert colloids also become trapped at the interfaces, suggesting that the observed attachment can be attributed to capillarity. Over time, these structures coarsen and eventually degrade, illustrating the dynamic nature of these systems. This system lays the foundation for future studies of the interplay of physicochemical properties of the fluid interfaces and bulk phases and microbial responses they provoke to induce complex spatial organization, to study species which occupy distinct niches, and to optimize efficient microbial cross-feeding or protection from competitors.

## I. INTRODUCTION

Fluid interfaces provide excellent environments to locate and organize particles, as evidenced by the rich and rapidly evolving literature focusing on the interaction and assembly of particles at interfaces.[1–7] Pieranski's seminal work demonstrated the trapping of colloids at interfaces and their organization into phases and structures which can be modulated by tuning interparticle interactions.[8] This work has spawned numerous studies in which capillary interactions and packing constraints are exploited to induce organization of particles into two- and three-dimensional structures.[6,9,10] When colloids are present in systems of two immiscible fluids, they attach to the interface to lower the interfacial area, thereby lowering the total system energy. In the case of conventional oil-water systems, this interfacial energy is significant and leads to a high trapping energy.[8] Depending on the nature of interactions between the immiscible fluids and the concentration of colloids, such systems can be stabilized in a range of states ranging from Pickering emulsions[11] and foams[12], to non-spherical droplets stabilized by particle monolayers.[13,14] Furthermore, biologically derived materials, bacteria, viruses, and spores have been exploited as "green" materials for stabilization of oil-water Pickering emulsions.[15] However, maintaining microbe viability through processing is not easily achieved.

Interfaces with far lower interfacial tension, formed, for example, by spinodal decomposition, also trap particles at the interface, as has been exploited to form bicontinuous interfacially-jammed emulsion gels (bijels).[16,17] We are particularly interested in studying colloidal behaviour in aqueous-aqueous systems. Interactions within these systems can be tuned to achieve a wide range of phase transitions, including phase separated systems in which the two phases form via spinodal decomposition, while maintaining favourable conditions to sustain life.[18] The resulting interfaces have an interfacial tension, albeit one which is orders of magnitude lower than that in the more conventional oil-water systems.[19–21] Our interest stems from the potential for these systems to sustain living species, opening the door to creating three dimensional structures using living colloids such as bacteria cells.

The development of non-trivial structures to confine and organize living colloids can have significant impact not only from a materials viewpoint, but also from a fundamental perspective in the study of the dynamics of living colloids in confinement. A rich literature is developing on the collective dynamics and self-organization of bacteria swimming in confined systems, including quasi-2D systems[22], 3D confining systems[23], and for bacteria trapped at or near fluid interfaces.[24–26] Aqueous two phase systems (ATPS), such as those studied here, can allow us to create soft, topographically complex or, in the case of bicontinuous systems or systems with interconnecting pathways, topologically complex systems in which to confine bacteria to study their dynamics. ATPS have been historically exploited for partitioning of biologics and are convenient to use due to the lack of a potentially toxic oil phase.[27] These prior works emphasize partitioning between phases, whereas here, we study assembly at the interfaces between the phases.

In this work, we have exploited ATPS to trap living bacteria at the interface, specifically, between casein-rich and xanthan-rich phases. Our results, for both living and inert colloids, strongly indicate that these bacteria attach to the interface to lower the interfacial area. These experiments serve as a proof of principle for this new domain of materials which can be used as dynamic, phase separating vehicles for studies in food science, medicine and materials science.

## II. RESULTS AND DISCUSSION

*Living colloids at the interface*

We exploit the aqueous phase separating system of casein and xanthan, which can be tuned by varying the relative concentrations of each component.[18,28] This system is relatively well studied, as it is widely exploited in the formulation of foodstuffs such as dairy products.[29] It is particularly interesting as a model system to investigate the interactions between living colloids and interfaces because these components are known to undergo spinodal decomposition under certain conditions, in particular when the casein is a part of skim milk powder (SMP).[18,30] We are working with solely casein and xanthan at concentrations near the previously reported spinodal point; Rhodamine B dye was added to the mixture to directly visualize the casein phase. Under these conditions, the system visually resembles an interconnected, bicontinuous structure. Furthermore, the protein phase has the potential to serve as a nutrient source for the bacteria in the system if the bacteria possess the correct enzymes, which allows the possibility for the bacteria to degrade and consume the structure over time.

We chose a wild type (WT) *E. coli* [MG1655] stained with SYTO 9 to ascertain whether living colloids attach to the interface of this phase separating casein-xanthan mixture. WT *E. coli* serves as an ideal model bacterium, as it has been widely manipulated and exploited in the study of recombinant DNA technology and as a host for production of recombinant proteins such as insulin.[31] Initially, as shown in Fig. 1A, the bacteria are scarce and are only observed swimming in the casein-void regions, indicating that they are not immediately driven to the interface (Movie S1). Following the progression shown in Fig. 1B – D, the bacteria attach to the interface within 16 h. Their accumulation at interfaces continues as the phase separation of the casein-xanthan mixture progresses; after 40 h, they have filled a large portion of the casein-xanthan interface. Fig. 1F, a projection of the compiled 3-D scan taken with a confocal laser scanning microscope (CLSM), gives clear evidence of bacteria at the interface of the interconnected casein-xanthan phases. By allowing the system to continue to evolve over one week post-mixing, we can see in Fig. 1E that there is barely any detectable red fluorescence remaining in the solution, indicating a loss of localization of the casein phase. Since the samples are kept at room temperature, the growth dynamics of the bacteria are slower than their typical growth at 37 ˚C, which allows our system to evolve and the bacteria to interact with its changing environment. At the end of one week, there were still live bacteria, as verified by plating a dilution of the contained mixture on 1.5 wt% agar gel plates made with LB broth, a standard method to assess colony growth.[32] The progression from Fig. 1A to Fig. 1E indicates the live bacteria are actively interacting with the matrix while being sequestered in the casein phase. As a control, the SYTO 9 dye was added to the casein-xanthan system devoid of bacteria; no increase or decrease in green or red fluorescence, respectively, was observed over time.

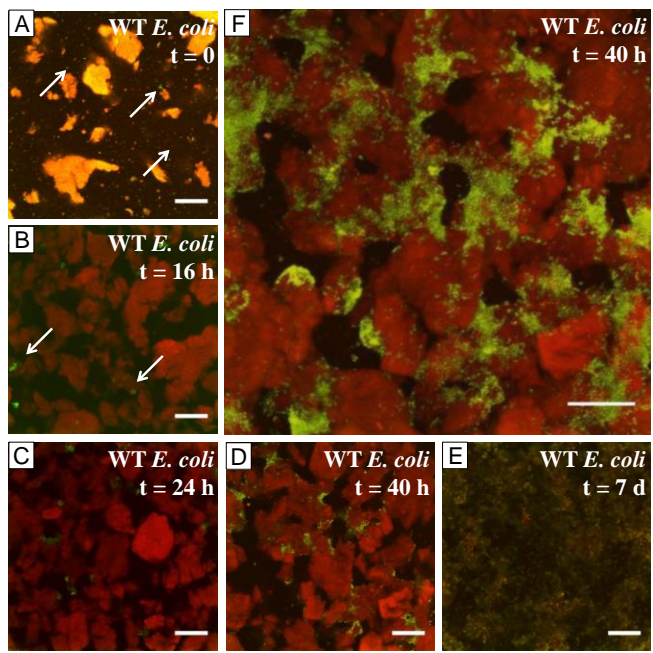

*Fig. 1. WT E. coli in casein-xanthan. Casein is stained red with Rhodamine B and E. coli is stained in green with SYTO 9. (A) Immediately after mixing, arrows showing free bacteria. (B) 16 h after mixing. (C) 24 h after mixing. (D) 40 h after mixing. (E) 7 d after mixing. (A-E) 20 μm above cover slip. (F) Projection of scan from top, 40 h after mixing. All scale bars 50 μm.*

An important question arises regarding the mechanism for the localization of the bacteria at the interface. Since the bacteria may consume casein, it is plausible that bacteria may swim toward the casein-rich interface via chemotaxis.[33] To test whether such cell motility is critical in inducing the attachment of bacteria to the interface between casein-rich and xanthan-rich phases, we use non-motile GFP *E. coli* [MDG10], which lack flagella, as our living colloids in the same phase-separating system. This strain also produces green fluorescent protein (GFP), and is therefore easily visualized by fluorescence microscopy. As shown in Fig. 2, these bacteria attach to the interface within 24 h, and the density of interface-trapped bacteria increases with time. When the system was allowed to evolve for longer times, in addition to densifying, the casein-rich region fluorescence again diminished as in the motile WT *E. coli* case.

We confirm that these non-motile GFP *E. coli* can survive on the casein by culturing them in a casein-added PBS solution, where the casein serves as the only possible carbon and nitrogen sources. As seen in Figure 2E, the bacteria have a slight increase in growth in the initial hours in the casein solution and are able to maintain a relatively constant number over a week. In contrast, the bacteria are not able to survive on the xanthan alone. Thus, once bacteria are trapped at the interface, their ability to slightly degrade the casein likely contributes to their survival and the subsequent long-time restructuring of the matrix.

We also observe interfacial attachment of other bacteria, including *Pseudomonas aeruginosa* (PA01), and *Pseudomonas sp.62* in this casein-xanthan system. For both species, the bacteria become trapped at the interface after ~16 h, and the phase-separated mixture coarsens. A key difference is that *P. sp. 62* and PA01 can actively break down and consume casein at higher rates than *E. coli*, so the casein region diminishes more rapidly (See SI). These results clearly demonstrate that phase separating aqueous mixtures can induce attachment of living colloids and sustain life, and that choosing specific bacterial strains affects the lifetime and dynamics of the matrix.

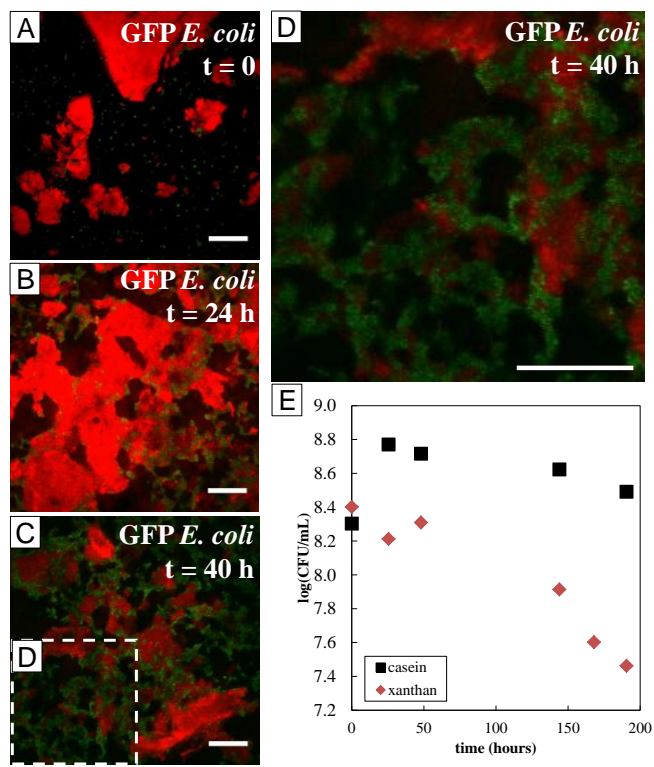

*Fig. 2. Non-motile, GFP E. coli in casein-xanthan mixture. Casein is stained red by rhodamine B and the E. coli is green. (A) Immediately after mixing. (B) 24 h after mixing. (C) 48 h after mixing. (D) Inset of (C). (E) Growth curve of GFP E. coli in 2 wt% casein solution (■) and 0.08 wt% xanthan (♦) in PBS, at 22 °C, measured by plating serially diluted bacterial suspensions from different time points on 1.5 wt% agar gel containing LB broth and counting colony forming units (CFU).[32] (A-D) taken 20 μm above cover slip, scale bar 50 μm.*

## *Inert colloids at the interface*

Our results based on motile and non-motile *E.coli* suggest that motility is not necessary to induce the attachment to the interface. To ascertain whether capillarity could be a likely mechanism, we estimate trapping energies owing to capillarity, *i.e.*, the energy associated with the reduction in the interfacial energy owing to particle attachment. The trapping energy of a colloid particle from an interface ($\Delta E$) can be estimated as $\Delta E \sim \gamma A$, where $\gamma$ and $A$ denote the interfacial tension and the area of interface eliminated by particle attachment, respectively. Reports have shown that interfacial tension between two aqueous phases made of caseinate and alginate, dextran and gelatin, or poly(ethylene glycol) and dextran can range from $10^{-3}$ to $10^{-2}$ mJ/m$^2$.[19–21] Using this representative value along with the typical size of bacteria (~ 1 μm), we estimate that the trapping energy at such an interface can exceed $10^3 k_B T$; this value is significant and indicates that once bacteria attach to the interface it may be extremely difficult to detach themselves from the interface. If, as this estimate suggests, capillary forces can account for the interfacial trapping of bacteria, then similar behaviour should be observable for inert colloids.

To confirm that the casein-xanthan system possesses sufficient interfacial energy to trap inert colloids at the interface, we added 0.01% green fluorescent polystyrene (PS) beads (*i.e.*, inert colloids) to the mixture at t = 0, Fig. 3A. After 24 h (Fig. 3B), the beads are found only at the interface between the casein and xanthan regions. After 40 h, shown in Fig. 3C, the phase separation has continued to progress and most of the casein regions within the viewing window are connected. This phase separation progression has occurred at room temperature; and is nearly bicontinuous after 40 h as shown in Fig. 3E. After this time point, the casein continues to densify and all components settle to the bottom. During the transition, however, we can see the inert colloids indeed remain trapped at the interface. Similar results have been reported recently for inert colloids in a phase separated system of gelatin and maltodextrin; the inert colloids were indeed trapped at the interface in a manner similar to oil-water-colloid three component systems.[34]

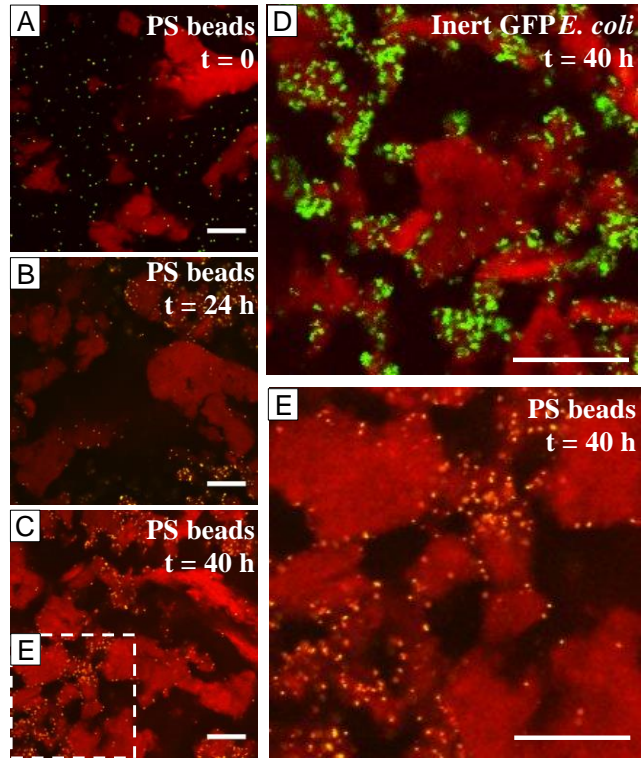

*Fig. 3. Inert colloids in casein-xanthan phase-separating system imaged by CLSM. Casein is stained red by Rhodamine B, green are green fluorescent polystyrene (PS) beads (A,B,C,E) or antibiotic-killed GFP E. coli (D). (Cells were killed with 200 µg/ml spectinomycin.) (A) PS beads immediately after mixing. (B) PS beads 24 h after mixing. (C) PS beads 40 h after mixing. (D) Dead GFP E. coli 40 h after mixing. (E) Inset of (C). All slices from 20 µm above glass, scale bars 50 µm.*

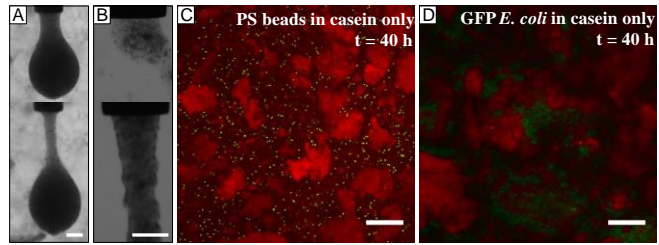

*Fig. 4. (A) Sequence of pendant drop experiment, drop fluid is 10 wt% casein, bulk fluid is 0.1 wt% xanthan, both filtered with 5 µm filter, scale bar 500 µm. (B) Sequence of pendant drop experiment, drop fluid is 10 wt% casein; bulk fluid is DI water, scale bar 500 µm. (C) PS beads in 2 wt% casein solution (no xanthan) 40 h after mixing. (D) GFP E. coli in 2 wt% casein solution (no xanthan) 40 h after mixing. (C) and (D) taken 20 µm above cover slip; casein is red, colloids are green; scale bar 50 µm.*

To complement the PS bead study with colloids that more closely match the shape and surface chemistry of live *E. coli*, we repeated this experiment with antibiotic-killed GFP *E. coli*, with similar results. Within 24 h, the dead bacteria were trapped at the interface and remained so as the system evolved, as seen in Fig. 3D. These results demonstrate that even weak interfacial tension can exert significant trapping energies, leading to attachment and trapping of particles at interfaces between two aqueous phases. The prolonged time required for the colloids to find the interface may be attributed to the highly viscous nature of the xanthan phase. The colloids have significantly dampened Brownian motion in this two phase system.

Xanthan is required to produce an interface with sufficient surface energy to capture microbes or colloids, as the following control experiments demonstrate. When a casein solution (10 wt%) was injected slowly into a xanthan solution (0.1 wt%) through a tube, we observed the formation of a transient drop which elongated continuously as shown in Fig. 4A. In contrast, when xanthan was removed from the continuous phase, it was not possible to form a cohesive drop no matter how slowly we injected the casein solution, as shown in Fig. 4B. These observations indicate that the presence of the casein-xanthan system leads to the development of interfaces that have a finite interfacial tension. Our conclusion regarding the importance of surface tension in the interfacial attachment of living and inert colloids is further corroborated by adding bacteria to casein-only solution. Fig. 4C and D demonstrate that in the absence of xanthan, living and inert colloids do not attach to the blobs of casein phase even after 40 h post mixing, indicating that the xanthan-casein interfacial tension, albeit extremely small, is critical for attachment to the interface.

The complex rheological nature of the xanthan-casein system prevents us from determining an exact value for this tension; the elasticity of the bulk phases plays a significant role in determining the shape of the droplet. This is particularly evident in the movies of drop elution (Movies S9-10), and in Fig. S3 in which one side of the droplet is torn away. Even after continued elution, the torn drop edge remains jagged, a form that is not consistent with an interface dominated solely by interfacial tension. These effects preclude determination of interfacial tension by drop shape methods, including spinning drop or pendant drop analyses.

### III.    CONCLUSIONS

We have demonstrated that inert and living colloids, including motile and immotile bacteria, attach to the interface between two aqueous phases. Our results strongly indicate that the attachment of these colloids is due to the presence of appreciable capillary trapping energies owing to the weak but finite interfacial tension between the two aqueous phases. This system, which uses only biologically compatible components, has potential for the development of novel biomaterials and bioreactors that take advantage of bacterial interfacial assemblies. Furthermore, these results can be recapitulated in other aqueous-aqueous systems, (e.g. PEG-dextran), opening new possibilities to biologically-inspired materials research – an avenue currently limited by well-known oil-water systems which do not generally maintain microbe viability and water-water partitioning systems which lack interfacial assembly.[15,27]

The organization that we observe may also be an important mechanism underlying the development of spatially structured microbial communities in natural settings, such as mixed-species biofilms. Bacteria synthesize numerous extracellular macromolecules, including polypeptides, polysaccharides, and nucleic acids, that form a protective matrix; xanthan gum secreted by the soil bacterium *Xanthomonas campestris* is one example. Macromolecules produced by distinct species may phase separate and drive specific bacterial populations to the interfaces, while the bacteria in turn control their localization by modulating the macromolecules anchored to their surface or by transforming the surrounding matrix. Thus, the interplay between the physicochemical properties of the fluid interfaces and the microbes' response to their environment may lead to complex spatial ordering that enables species to occupy distinct niches and optimize efficient cross-feeding or protection from competitors.

## IV. MATERIALS AND METHODS

*Stock Solution Preparation*

Stock casein and xanthan were prepared following the procedures described in the numerous phase separation studies.[18,28,30] The casein, obtained from Fisher, was used to prepare a 10 wt% solution; the casein was stirred with 0.1 wt% NaOH in water at 55°C for 2 hours then stirred at room temperature for another 2 hours. The stock xanthan, generously supplied by CP Kelco, (Keltrol – food grade xanthan gum) was used to make a 0.2 wt% solution by mixing 0.1 M NaCl with the xanthan in water and stirring at 70°C for one hour before stirring overnight at room temperature. Both stocks were then kept at 4°C.

*Sample Cell Preparation*

The solutions were placed in a well formed within a polydimethylsiloxane (PDMS) slab by using a rectangular mold, suitable for placement on a microscope stage. The well holds 50 μL of solution, and is approximately 500 μm deep. It was filled with the solution to be observed, and immediately sealed using a coverslip as a lid, as depicted in Fig. S1. When the wells were not being imaged, they were kept in a 100% humidity chamber to prevent evaporation, and wrapped in a protective foil to prevent photobleaching of fluorophores used to label various constituents.

*Bacteria Preparation*

Four species of bacteria were used in this study: *Pseudomonas aeruginosa* (PA01), *Pseudomonas sp.62*, non-motile GFP-labelled *Escherichia coli* (MDG10 – a GFP-labelled derivative of the non-motile and flagella-minus *Escherichia coli* strain MC4100[35,36]), and motile WT *E. coli* (MG1655). All species were cultured at room temperature in LB broth for 2 days and then stored at 4°C for 1 day. These bacterial suspensions were then centrifuged and the pellet was re-suspended in phosphate buffered saline (PBS) before adding to the sample solution.

*Solution Preparation*

The solutions and suspensions used in experiments were mixed approximately 20 min before *t = 0*. In all studies, the casein concentration was kept constant at 2 wt%. Xanthan was added at 0.08 wt% and the bacterial suspension was added at about 0.01% solids. PBS was added to achieve the desired wt% after addition of the stock solutions. 5 μL of a 0.05 wt% Rhodamine B (Fisher) solution was added per 1 g sample. Green fluorescent polystyrene (PS) beads (1 μm diameter) from Sigma were used for the non-bacteria experiment, also at 0.01% solids. For the non-fluorescent bacteria, a 3 μL/100 mL SYTO 9 solution was also added at 50 μL/1 g solution. The solution was mixed on the vortex mixer for 3 pulses and added to the PDMS cell. The final pH of the solution was approximately 6.3.

*Sample Imaging*

Solutions were imaged using confocal laser scanning microscopy (CLSM) on an Olympus IX81. The 512 x 512 pixel images were taken using a 40X water immersion lens excited by 488 nm and 543 nm lasers and collected through EGFP and Rhod 2 filter sets. Images were compiled using ImageJ Software.

*Acknowledgements*

This work is supported by the National Science Foundation through PENN MRSEC DMR-1120901. S.D.H. was supported by a fellowship through GAANN grant P200A120246.

*Notes and references*

*a* Department of Chemical and Biomolecular Engineering
  University of Pennsylvania, Philadelphia, PA 19104, USA

*b* Department of Biology
  University of Pennsylvania, Philadelphia, PA 19104

*Co-corresponding authors:
  D.L. (email) daeyeon@seas.upenn.edu
  K.J.S. (email) kstebe@seas.upenn.edu

**SUPPLEMENTARY INFORMATION**

*Experimental Setup.* PDMS wells were created with ~500 μm deep molds. The solution was placed in the well and closed with a coverslip. The device was turned over to image on an inverted confocal laser scanning microscope, as depicted in Fig. S1.

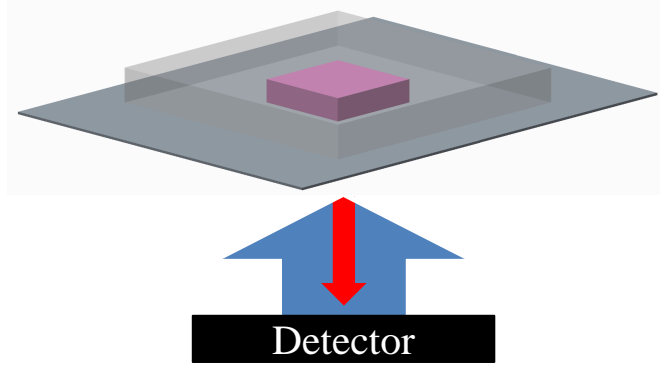

*Fig. S1. Schematic of experimental setup. Casein-xanthan suspension was enclosed in a PDMS cell on top of a coverslip. Total sample thickness was 500 μm. Imaged on inverted confocal laser scanning microscope (CLSM).*

*Motility study.* We monitored both WT and GFP *E. coli* strains during various periods throughout the phase separation at 20 μm above the coverslip for short periods of time. These movies (S1-S3) reiterate the free motion that the WT *E. coli* has immediately after mixing versus 24 hours after mixing. Once trapped at the interface, they do not swim. This lack of motility is either because they are trapped at the interface and cannot move or they have turned off their motility gene. No additional verification was done investigating the gene expression through time. The lack of motion of the GFP *E. coli* in Movie S3 is shown as verification that we have used a non-motile strain.

| | |
|---|---|
| MovieS1.avi | Movie S1. WT *E. coli* within sample well are monitored 20 μm from cover slip immediately after mixing. Casein is stained red and the *E. coli* are stained green with SYTO 9. The bacteria can be observed swimming across the viewing window multiple times. Scale bar 50 μm. |
| MovieS2.avi | Movie S2. WT *E. coli* within sample well are monitored 20 μm from cover slip 24 h after mixing. Casein is stained red and the *E. coli* are stained green with SYTO 9. The bacteria are no longer observed swimming across the viewing window. Scale bar 50 μm. |
| MovieS3.avi | Movie S3. Non-motile GFP *E. coli* within sample well are monitored 20 μm from cover slip immediately after mixing. Casein is stained red and the *E. coli* are green. The bacteria are not moving, nor are they at the interface yet. Scale bar 50 μm. |

*CLSM scans of WT E.coli, GFP E. coli, and inert colloids.* Representative Z-scans of the slices shown in Fig. 2 – 4 are given below in Movie S4-S8. In each movie, as in the main paper, red indicates the casein region, stained with Rhodamine B, green indicates live, WT *E. coli*. All scale bars are 50 μm.

| 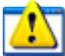 MovieS4.avi | Movie S4. Z-scan from top to coverslip of live, WT *E. coli*, t = 0 |
|---|---|
| 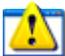 MovieS5.avi | Movie S5. Z-scan from top to coverslip of live, WT *E. coli*, t = 16 h |
| 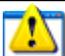 MovieS6.avi | Movie S6. Z-scan from top to coverslip of live, WT *E. coli*, t = 24 h |
| 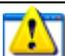 MovieS7.avi | Movie S7. Z-scan from top to coverslip of live, WT *E. coli*, t = 40 h |
| 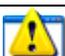 MovieS8.avi | Movie S8. Z-scan from top to coverslip of live, WT *E. coli*, t = 7 d |

*CLSM scans of Pseudomonas strains.* Two *Pseudomonas* strains were added to the system to see the behavior. *Pseudomonas aeruginosa* (PA01) studies were carried out exactly as the *E. coli* studies grown in a hydrated, light-protected environment at room temperature (Fig. S1). The images presented in Fig. S2 are mixtures of 2 wt% casein in PBS (no xanthan). Including xanthan drives the bacteria to the interface, as shown in the main paper, and the PA01 decomposes the casein region within 24 hours. It is easier to see the breakdown of the casein region without the xanthan, as seen in Fig. S2B, in which the red fluoresced region appears patchy with bacteria around these regions. PA01 was stained green with SYTO 9 stain. PA01 is a positive casein clearer so it follows that the casein region gets cleared more rapidly with this bacterium compared to *E. coli*. As such, after only 40 h (Fig. S1C), the PA01 sample resembles the 7 d *E. coli* sample (Fig. 1E) with its lack of red fluorescence and apparent flattening. The species *P. sp. 62* is a marine bacteria that grows ideally around 27 °C and a representative figure is presented in Fig. S3. The bacteria attach at the casein-xanthan interface after 24 h but interestingly do not consume the matrix as rapidly as the PA01.

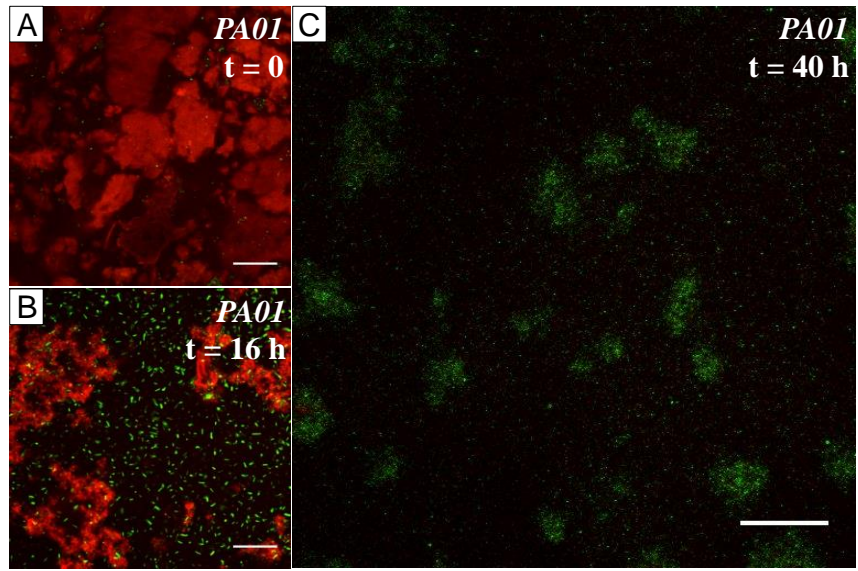

*Fig. S2. Time progression of P. aeruginosa (PA01) imaged in CLSM 20 μm above coverslip. Casein is stained with Rhodamine B and PA01 is stained with SYTO 9. (A) Immediately after mixing (B) 16 hours after mixing (C) 40 hours after mixing. All scale bars are 50 μm.*

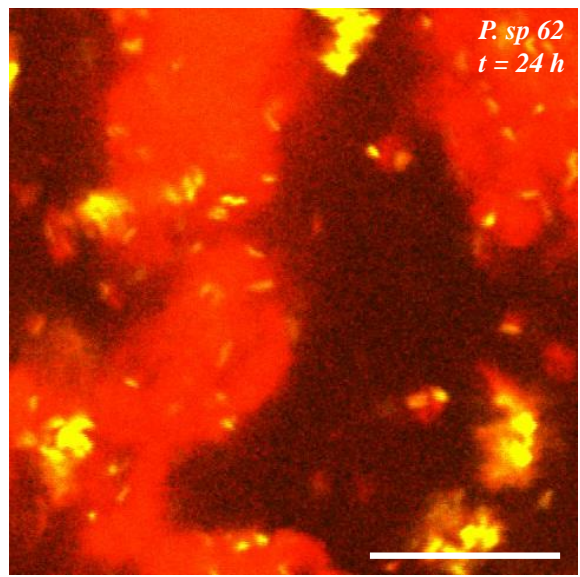

*Fig. S3. Image of Pseudomonas sp. 62 in 2 wt% casein, 0.08 wt% xanthan solution 20 μm above coverslip. Scale bar 20 μm.*

***Pendant drop studies***. Simple pendant drop experiments were performed to qualitatively characterize the interfacial tension between the casein and xanthan phases as shown in the main text in Fig. 5A and 5B. Full drop movies shown in Movie S9 and S10. Non-Newtonian characteristics are especially evident in Fig. S4, in which the drop has a mostly uniform shape but of which part is torn.

| | Movie S9. 10 wt% casein solution eluted into 0.1 wt% xanthan solution. Corresponds to stills from Fig. 5A. |
|---|---|
| MovieS9.avi | |

| 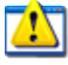 MovieS10.avi | Movie S10. 10 wt% casein solution eluted into DI water. Corresponds to stills from Fig. 5B. |
|---|---|

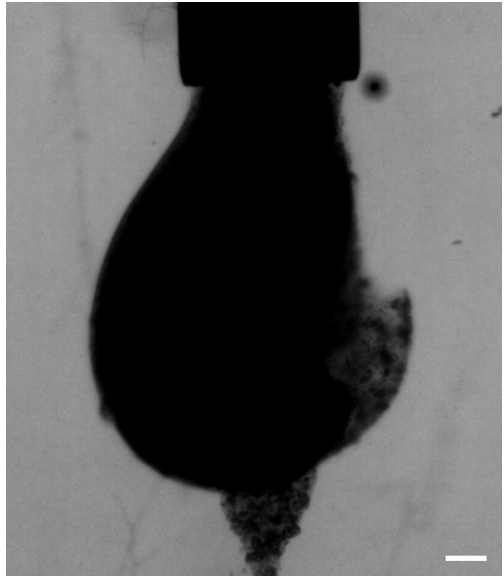

*Fig. S4. Frame from pendant drop experiment in which 10 wt% casein was eluted into 0.1 wt% xanthan. Scale bar is 2000 μm.*